\documentclass[
aps,prd,
preprintnumbers,
amsmath,
amssymb,
nofootinbib,
preprintnumbers,
longbibliography
]{revtex4-1}
\usepackage{here}
\usepackage{footnote}
\usepackage{comment,braket}
\usepackage{epsf}
\usepackage{amsmath}
\usepackage{graphics}
\usepackage{amsfonts}
\usepackage{amssymb}
\usepackage{latexsym}
\usepackage{color}
\usepackage{natbib}
\usepackage{graphicx}
\usepackage{hyperref}
\usepackage{array}

\usepackage[svgnames]{xcolor}
\definecolor{phthaloblue}{rgb}{0.0, 0.06, 0.54}
\hypersetup{
    colorlinks=true,
    linkcolor=blue,
    citecolor=blue,
    filecolor=blue,
    urlcolor=blue,
    }

\begin{document}
\title{Thermal Ringdown of a Kerr Black Hole:\\Overtone Excitation, Fermi-Dirac Statistics and Greybody Factor}

\author{Naritaka Oshita}
\email{naritaka.oshita@riken.jp}
\preprint{RIKEN-iTHEMS-Report-22}
\affiliation{
  $^1$RIKEN iTHEMS, Wako, Saitama, Japan, 351-0198
}

\begin{abstract}
We find a significant destructive interference among Kerr overtones in the early ringdown induced by an extreme mass-ratio merger of a massive black hole and a compact object, and that the ringdown spectrum apparently follows the Fermi-Dirac distribution. We numerically compute the spectral amplitude of gravitational waves induced by a particle plunging into a Kerr black hole and study the excitation of multiple quasi-normal (QN) modes. We find that the start time of ringdown is before the strain peak of the signal and corresponds to the time when the particle passes the photon sphere. When the black hole has the near-extremal rotation, the Kerr QN frequencies are close to the fermionic Matsubara frequencies with the Hawking temperature and the chemical potential of the superradiant frequency. We indeed find that the absolute square of the spectral amplitude apparently follows the Fermi-Dirac distribution with the chemical potential of around the real QN frequency of the fundamental mode. Fitting the Boltzmann distribution to the data in higher frequencies, the best-fit temperature is found out to be close to the Hawking temperature, especially for rapid rotations. In the near-extremal limit, the gravitational-wave spectrum exhibits a would-be Fermi degeneracy with the Fermi surface at the superradiant frequency $\omega = \mu_{\rm H}$. We show that the greybody factor, i.e., the absorption cross section of a black hole, leads to the Fermi-Dirac distribution. As the greybody factor is another no-hair quantity of black holes, this opens a new possibility that we can test general relativity by observationally searching for the Boltzmann distribution in $\omega \gtrsim \mu_{\rm H}$ without extracting QN modes from ringdown. We could measure the mass and angular momentum of ringing black holes and could probe the Kerr/CFT correspondence by measuring the greybody factor imprinted on the ringdown spectrum.
\end{abstract}

\maketitle

\section{Introduction}
A black hole is one of the simplest astrophysical objects in the Universe as it has only three {\it hairs} in general relativity, i.e., mass, angular momentum, and charge. The quasi-normal (QN) modes of a black hole are also characterized only by the three hairs by virtue of the no-hair theorem. A black hole ringing results in the emission of gravitational-wave (GW) ringdown, whose waveform is represented by a superposition of QN modes. Each QN mode is represented as a damped sinusoid and has a complex frequency, whose real and imaginary parts are the frequency and damping rate of the QN mode, respectively. The GW ringdown signal is emitted during the relaxation process of a ringing black hole. For instance, a binary black hole merger eventually leads to the emission of ringdown signal soon after the two progenitor black holes merge. To date, the detection of ringdown signals sourced by binary black hole merger events have been reported, e.g., in Refs. \cite{LIGOScientific:2016lio,Isi:2019aib,LIGOScientific:2020tif,Capano:2021etf}.

The poles of the retarded Green's function associated with the gravitational perturbations of a black hole are nothing but the QN modes of the black hole \cite{Leaver:1986gd,Sun:1988tz,Andersson:1995zk,Glampedakis:2001js,Glampedakis:2003dn,Nollert:1992ifk,Andersson:1996cm,Nollert:1998ys}. In the context of the holographic principle, it is conjectured that there would be a duality between the QN modes of a black hole and the poles of the retarded Green's function associated with the corresponding conformal field theory (CFT).
As a supporting evidence, it was shown that a Ba\~{n}ados-Teitelboim-Zanelli black hole has its QN modes dual to the poles of the thermal Green's function in the corresponding CFT \cite{Birmingham:2001pj}. Also, even Kerr black holes that exist in our Universe may have the holographic nature according to the Kerr/CFT correspondence \cite{Guica:2008mu} (see also \cite{Bredberg:2011hp,Compere:2012jk,Bonelli:2021uvf}). The near-horizon geometry of an extremal Kerr black hole, which has a quotient of AdS${}_3$ space at a fixed polar angle, can be mapped to a dual two-dimensional CFT. Indeed, it was reported that the scattering of a Kerr black hole agrees with the thermal CFT correlators \cite{Bredberg:2009pv,Cvetic:2009jn,Hartman:2009nz,Chen:2010ni}. Also, the hidden conformal symmetry associated with the photon sphere was recently proposed in \cite{Raffaelli:2021gzh}. Those supporting evidence for the holographic nature of black holes imply that the geometrical degrees of freedom in the vicinity of a black hole would correspond to a lower-dimensional CFT. As such, it is highly motivated to study the perturbation of a Kerr black hole in the context of the holographic principle.

Besides the motivation arising from the holography, the detection of excited multiple QN modes is important to test general relativity in strong-gravity regime. Recently, the excitation of multiple QN modes was confirmed around the strain peak of GW signal \cite{Giesler:2019uxc,Ma:2021znq,Li:2021wgz}, at least based on the fit of QN modes to the numerical relativity waveforms \cite{Boyle:2019kee}. The fit of the QN modes to the GW data of GW150914 was also performed and the detection of the first overtone was claimed \cite{Isi:2019aib} although it may be still controversial\footnote{For the discussion on the start of ringdown based on the precession of the remnant, see Ref. \cite{Hamilton:2021pkf}.} as was discussed in Ref. \cite{Cotesta:2022pci,Isi:2022mhy}. If ringdown can start at around the strain-peak time, the multiple QN modes can be detected with a higher signal-to-noise ratio. As such, the excitation of multiple QN modes, including overtones, has been actively studied in various contexts (see, e.g., Refs. \cite{Bhagwat:2019dtm,Bhagwat:2019bwv,Cook:2020otn,JimenezForteza:2020cve,Capano:2020dix,Bustillo:2020buq,Dhani:2020nik,Mourier:2020mwa,Finch:2021iip,Isi:2021iql,Forteza:2021wfq,Dhani:2021vac,Ota:2021ypb,Sago:2021gbq,Sberna:2021eui,Konoplya:2022hll,Ma:2022wpv}). Still, the ringdown emission or significant excitation of overtones at earlier times can be controversial. Perhaps, one of the most crucial issues is the linearlity problem of the black hole ringing. For instance, in the head-on collision of two non-spinning black holes, the two holes do not completely merge immediately, but the initial two marginally outer trapped surfaces (MOTSs) continue to exist and an outermost common MOTS forms and behind which the highly non-linear region exists \cite{Pook-Kolb:2019iao,Pook-Kolb:2019ssg}. Only the outside of the outermost MOTS may be relevant to what we observe at a distant region, provided that the MOTS has an analog in the event horizon. Therefore, if these are the cases even for a collision of two rotating black holes and if the geometry {\it outside} the MOTS can be described by the Kerr spacetime with linear perturbations, we could observe the GW ringdown signal regardless of the non-linearity inside the MOTS. A similar idea was carefully probed and discussed based on the numerical relativity approach for a non-rotating system \cite{Mourier:2020mwa}.

In this paper, we show observable supporting evidence for the holographic nature of a Kerr black hole. We argue that the excitation of mulptiple QN modes would follow the Fermi-Dirac statistics, at least for an extreme mass ratio merger. We numerically compute the spectral amplitude of a GW signal induced by an extreme mass ratio merger involving a massive black hole, with mass $M$ and angular momentum $J$, and a plunging small-mass compact object with mass $m_{\rm p}$. Our computation methodology is based on Ref. \cite{Kojima:1984cj}, where the self force of the particle is ignored and the whole signal is computed in a linear manner, that is, the background spacetime is fixed. Therefore, any non-linear effects are not involved in our analysis. This picture is valid only for an extreme mass ratio $M \gg m_{\rm p}$. Fitting multiple QN modes of the Kerr black hole to the obtained spectral data in the frequency domain and performing the inverse Fourier transformation to obtain the best-fit waveform in the time domain, we find a significant destructive interference among overtones occurs at the beginning of ringdown, followed by the strain peak and exponential damping. Such a strong destructive interference is possible only when multiple QN modes are excited simultaneously. We then carefully analyze the obtained spectral data and find out that the absolute square of the spectral amplitude for frequencies higher than the real QN frequency of the fundamental mode can be modeled by the Boltzmann distribution. Identifying the superradiant frequency as a chemical potential, we arrive at the modeling of ringdown spectrum with the Fermi-Dirac distribution. Fitting the Boltzmann factor to the spectrum, we find the best-fit temperature takes a value close to the Hawking temperture (or Hawking {\it frequency}\footnote{The Hawking frequency, $T_{\rm H}$, takes the same value of the Hawking temperature given by $\hbar / k_{\rm B} \times T_{\rm H}$ in the natural unit of $\hbar = k_{\rm B} = 1$. We consider a classical process and the Hawking temperature including the Planck constant is not involved in our analysis. The Hawking ``temperature" we here consider has the dimension of ``frequency" and it is defined without the Planck constant.}), especially for rapid rotations. We will show that the higher a black hole rotation is, the more overtones are excited. These imply that the excitation of multiple overtones can be modeled by the Fermi-Dirac distribution.
Our finding is consistent with the analogy between the QN modes for a rapidly rotating black hole and the poles of the Fermi-Dirac distribution, known as the Matsubara modes. Also, the analogy between the ringdown spectrum and the Fermi-Dirac distribution opens a novel possibility such as a modeling of ringdown waveforms with the Fermi-Dirac distribution or with the superposition of the Matsubara modes. Also, searching for the Boltzmann distribution in an observed GW spectrum, including a ringdown signal, from a rapidly rotating black hole\footnote{For a review of rapidly rotating, i.e., near-extremal, supermassive black holes, see e.g. Ref. \cite{Reynolds:2020jwt}.} may put the thermal or holographic picture of Kerr black holes to the test.

The paper is organized as follows. In the next section, we briefly review the procedure to compute a GW signal induced by a particle plunging into a Kerr black hole. In Sec. \ref{sec_destructive_interference}, by fitting QN modes to numerically obtained data, we verify the excitation of overtones and search for the best-fit start time of ringdown. We then show that a significant destructive interference occurs at the beginning of ringdown, which is possible only when multiple overtones are excited simultaneously. In Sec. \ref{sec_thermal_overtones}, we first show that the greybody factor is imprinted on the ringdown in an analytic manner. Then, motivated by that the frequency dependence of the greybody factor is similar to that of the Fermi-Dirac distribution\footnote{The greybody factor is usually defined as the transmissivity of the angular momentum barrier in the context of the Hawking radiation. In this paper, however, we referred the greybody factor as the reflectivity of the barrier since the reflectivity is simply related to the transmissivity via $ (\text{reflectivity}) = 1 - (\text{transmissivity})$.}, we propose the {\it thermal ringdown model} for which the spectrum of GW ringdown is modeled by the Fermi-Dirac distribution. We indeed find the Fermi-Dirac-like distribution in the simulated GW spectrum and obtain the best-fit temperature associated with the Boltzmann factor at higher frequencies in the spectrum. We find that it takes a value close to the Hawking temperature. We will also show that in the near-extremal situation, for which the Hawking temperature is almost zero, the GW spectrum exhibits a would-be Fermi degeneracy with the Fermi surface at the superradiant frequency. We also investigate the modeling of ringdown with the fermionic Matsubara modes. 
Our conclusions and discussions are provided in Sec. \ref{sec_conclusions}.

\section{gravitational wave emission induced by a falling particle}
\label{sec_methodology}
In this section, we compute waveforms of GW induced by a particle with mass $m_{\rm p}$ plunging into a Kerr black hole, whose mass and angular momentum are denoted by $M$ and $J$, respectively. For an extreme mass ratio, $M \gg m_{\rm p}$, the particle does not non-linearly disturb the background spacetime, and one can compute the GW signal in the linear approximation. In this case, a signal is obtained by solving the Teukolsky equation \cite{Teukolsky:1973ha}. In this manuscript, we solve the Sasaki-Nakamura (SN) equation \cite{Sasaki:1981sx} that has a short-range angular momentum barrier and is, therefore, easier to solve numerically. The source term determined by the dynamics of the falling particle is given in Ref. \cite{Kojima:1984cj}. The line element on the Kerr spacetime has the form
\begin{equation}
ds^2 = - \left( 1-\frac{2Mr}{\Sigma} \right) dt^2 + \frac{\Sigma}{\Delta} dr^2 + \Sigma d\theta^2 + \sin^2 \theta \left( r^2+a^2 + \frac{2Mra^2 \sin^2 \theta}{\Sigma} \right) d \varphi^2 - \frac{4 Mra\sin^2 \theta}{\Sigma} dt d\varphi,
\end{equation}
where $a\equiv J/M$ and the functions, $\Sigma$ and $\Delta$, are defined as
\begin{align}
\Sigma &\equiv r^2 + a^2 \cos^2 \theta,\\
\Delta &\equiv r^2 -2M r + a^2 = (r-r_+) (r-r_-),\\
\text{with} \ r_{\pm} &\equiv M\pm \sqrt{M^2 -a^2}.
\end{align}
The SN equation with the SN perturbation variable, $X_{lm} = X_{lm} (\omega, r^{\ast})$, is given by
\begin{equation}
\left(\frac{d^2}{dr^{\ast} {}^2} - F_{lm} \frac{d}{dr^{\ast}} - U_{lm} \right) X_{lm} = \tilde{T}_{lm},
\label{SNeq}
\end{equation}
where $r^{\ast}$ is the tortoise coordinate defined by
\begin{equation}
r^{\ast} = \int dr \frac{r^2+a^2}{\Delta} = r+ \frac{2M}{r_+-r_-} \left[r_+ \ln{\left( \frac{r-r_+}{2M} \right)} -r_- \ln{\left( \frac{r-r_-}{2M} \right)} \right],
\end{equation}
the source term $\tilde{T}_{lm}$ is
\begin{equation}
\tilde{T}_{lm} = \frac{\gamma_{lm} \Delta W_{lm}}{(r^2+a^2)^{3/2} r^2} \exp{\left( -i \int \frac{(r^2+a^2)\omega -am}{\Delta} dr \right)},
\label{source_factor_SN}
\end{equation}
and the explicit forms of $F_{lm}$, $U_{lm}$, $\gamma_{lm}$, and $W_{lm}$ are shown in Appendix \ref{app_explicit_forms}. The label $(l,m)$ specifies an angular mode of the spin-weighted spheroidal harmonics ${}_{-2} S_{lm} (a\omega, \theta) e^{im \varphi}$. The function $W_{lm} = W_{lm}(r,\omega)$ is determined by the trajectory of an plunging particle, whose proper time is denoted by $\tau$. The trajectory on the equatorial plane $(t(\tau), r (\tau),\varphi (\tau), \pi/2)$ is obtained by solving these equations \cite{Carter:1968rr,Misner:1973prb,Kojima:1984cj}:
\begin{align}
r^2 \frac{dt}{d\tau} &= -a (a-L_{\rm p}) + \frac{r^2+a^2}{\Delta} P,\label{geot}\\
r^2 \frac{d\varphi}{d\tau} &=- (a-L_{\rm p}) + \frac{a}{\Delta} P,\label{geop}\\
r^2 \frac{dr}{d\tau} &= - \sqrt{Q},\label{geor}
\end{align}
where $L_{\rm p}$ is the orbital angular momentum of the particle and
\begin{align}
P & \equiv r^2+a^2 -L_{\rm p} a,\\
Q & \equiv 2 M r^3 -L_{\rm p}^2 r^2 + 2Mr (L_{\rm p} - a)^2.
\end{align}
Also, these geodesic equations correspond to the case of $E=m_{\rm p}$, where $E$ is the energy of the particle, and the zero of the Carter constant (see, e.g., Ref. \cite{Sago:2020avw}).
The SN equation can be solved with the boundary condition of
\begin{align}
X_{lm} (\omega, r^{\ast}) = 
\begin{cases}
X_{lm}^{\rm (out)} (\omega) e^{i \omega r^{\ast}} & \text{for} \ \ r^{\ast} \to + \infty,\\
X_{lm}^{\rm (in)} (\omega) e^{- i k r^{\ast}} & \text{for} \ \ r^{\ast} \to - \infty,
\end{cases}
\end{align}
where $k \equiv \omega - m\Omega_{\rm H}$ and $\Omega_{\rm H} \equiv a /(2Mr_+)$. The strain $h = h(t,r)$ is obtained from $X_{lm}^{\rm (out)}$ as 
\begin{align}
&h=\sum_{l,m} h_{lm} =\sum_{l,m} h_{+,lm} + i h_{\times, lm} = \sum_{l,m} \frac{e^{i m \varphi}}{\sqrt{2 \pi} r} \int_{- \infty}^{\infty} d \omega \tilde{h}_{lm}(\omega, \theta) e^{-i \omega (t-r^{\ast})},\\
&\tilde{h}_{lm} (\omega, \theta) \equiv - \frac{2}{\omega^2} {}_{-2} S_{lm} (a\omega, \theta) R_{lm} (\omega),\\
&R_{lm} (\omega) \equiv \frac{-4 \omega^2 X^{\rm (out)}_{lm}}{\lambda_{lm} (\lambda_{lm} + 2) -12 i M \omega -12 a^2 \omega^2},
\end{align}
where $\lambda_{lm}$ is the eigenvalue determined by the regularity of the spheroidal harmonics:
\begin{align}
S_{lm} (\omega, \theta) &\sim
\begin{cases}
(1+\cos \theta)^{k_+} \ \ &(\theta \to 0),\\
(1+\cos \theta)^{k_-} \ \ &(\theta \to \pi),
\end{cases}\\
&\text{with} \ k_{\pm} \equiv |m\mp 2|/2.
\end{align}
One can obtain $X_{lm}^{\rm (out)}$ with the Green's function technique as
\begin{equation}
X_{lm}^{\rm (out)} (\omega) = \int dr' \frac{\tilde{T}_{lm} (r', \omega) X_{lm}^{\rm (hom)} (r', \omega)}{2i\omega B_{lm}(\omega)},
\label{X_out}
\end{equation}
where $X_{lm}^{\rm (hom)}$ is a homogeneous solution to the SN equation, i.e., a solution of the SN equation with $\tilde{T}_{lm} = 0$, with the boundary condition of
\begin{align}
X_{lm}^{\rm (hom)} = 
\begin{cases}
A_{lm}(\omega) e^{i \omega r^{\ast}} + B_{lm}(\omega) e^{-i \omega r^{\ast}} \ \ &(r^{\ast} \to + \infty),\\
e^{-i k r^{\ast}} \ \ &(r^{\ast} \to -\infty).
\end{cases}
\end{align}
The coefficient $B_{lm} (\omega)$ is important since it has zeros at complex frequencies, $\omega = \omega_{lmn}$, which are the poles of the Green's function. They are also known as the QN modes of a Kerr black hole.
\begin{figure}[t]
  \begin{center}
    \includegraphics[keepaspectratio=true,height=60mm]{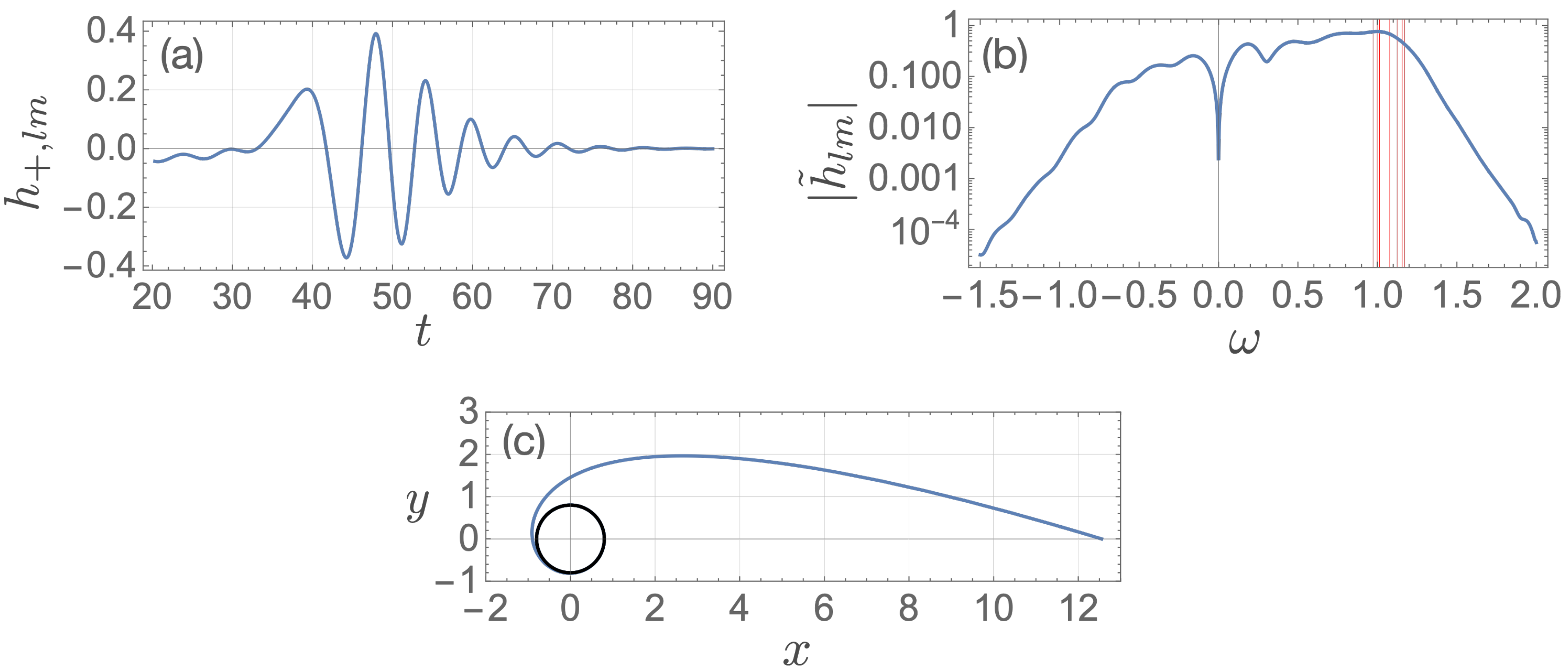}
  \end{center}
\caption{Plots of the GW signal with $j=0.8$ and $l=m=2$ (a) in the time domain and (b) in the frequency domain. Red lines in (b) indicate the values of $\text{Re} (\omega_{22n})$ with $n=0,1,...7$. The trajectory of the plunging particle with $j=0.8$ and $L_{\rm p}=1$ is shown in (c).
}
\label{lm22}
\end{figure}

Note that in this formalism, the backreaction on the falling particle is ignored. This approximation does not work at all when its orbital angular momentum $L_{\rm p}$ takes either the critical value of $-2M (1+\sqrt{1+j})$ or $2M (1+\sqrt{1-j})$, for which the particle takes infinite time to fall into the black hole \cite{Kojima:1984cj}. Here we define the non-dimensional spin parameter $j\equiv a/M$. Therefore, the value of $L_{\rm p}$ is limited to 
\begin{equation}
-2M (1+\sqrt{1+j}) < L_{\rm p} < 2M (1+\sqrt{1-j}).
\end{equation}

\begin{figure}[h]
  \begin{center}
    \includegraphics[keepaspectratio=true,height=38mm]{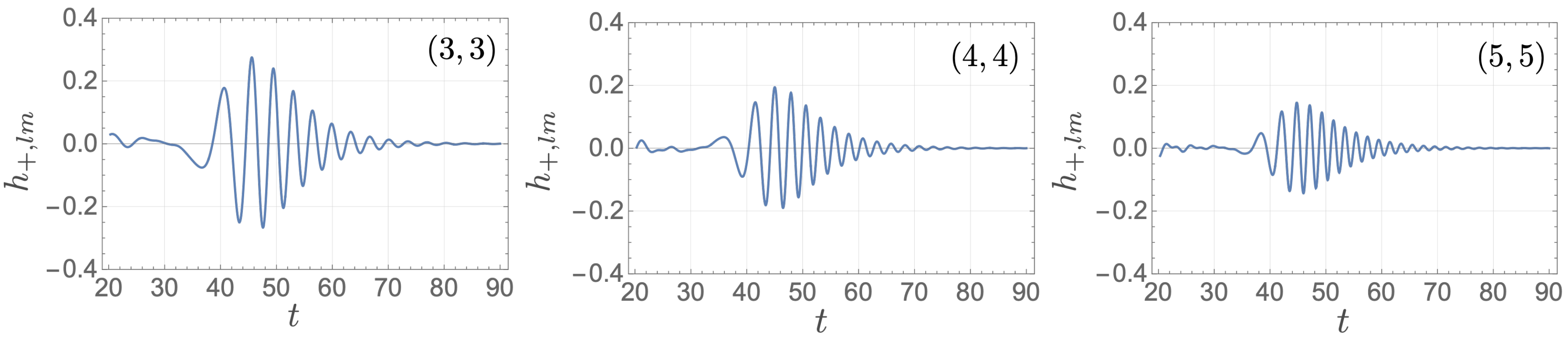}
  \end{center}
\caption{GW waveforms in the time domain for $(l,m) = (3,3)$, $(4,4)$, and $(5,5)$.
The spin parameter is $j=0.8$.}
\label{345}
\end{figure}

We numerically solve the geodesic equations (\ref{geot}-\ref{geor}) and homogeneous SN equation with the 4th-order Runge-Kutta method. The detailed discussion on the resolution of our numerical computation is provided in Appendix \ref{app_resolution}. The spectrum and time-domain waveform for $(l,m) = (2,2)$ are shown in Figure \ref{lm22} along with the trajectory of the particle. Throughout the manuscript, the parameter is set to $M = 1/2$ and the amplitude is normalized by a factor of $m_{\rm p}/r_{\rm obs}$, where $r_{\rm obs} (\gg M)$ is a radial position of an observer detecting the signal on the equatorial plane. One can see that the peak in frequency-domain comes around the real part of QN frequencies. In Figure \ref{345}, one can see that the GW signal of higher angular modes induced by the falling particle have smaller amplitudes compared to that of $(l,m) = (2,2)$. We fix the orbital angular momentum of the particle to $L_{\rm p} = 1$ throughout the manuscript, but the change of $L_{\rm p}$ does not affect the qualitative result presented in the following (see Appendix \ref{app_Lp}). In the following, we will often omit the subscript of $(l,m)$ for the sake of brevity. 

\section{Destructive Interference of Excited quasi-normal modes}
\label{sec_destructive_interference}
We fit QN modes to the obtained GW signal in the frequency domain to extract the start time of ringdown without truncating the original waveform data, including orbital and ringdown signals. We then show that the start time of ringdown is before the strain-peak time, where a destructive interference among multiple overtones occurs.
Fitting higher overtones to a ringdown, beginning before the strain peak, in time domain may be difficult as each higher overtone is exponentially enhanced at earlier times, which makes the fitting in time domain unstable. The fitting analysis in the frequency domain has the advantage of fitting QN modes in a stable way even for a GW signal whose rigndown starts before the strain peak.

Let us begin with the modeling of GW ringdown waveforms in the frequency domain. In the GW spectrum, $\tilde{h} (\omega)$, computed in the previous section, we have two main signals, orbital and rindown signals, and it is obvious that the orbital signal is emitted, and the ringdown emission follows. The start time of ringdown, $t^{\ast}$, is encoded in the spectral amplitude, $\tilde{h}(\omega)$, with the factor of $e^{i \omega t^{\ast}}$. To demonstrate this, let us first introduce the ringdown-waveform model in the time domin:
\begin{equation}
h_{\rm QNM} \equiv \sum_{lm} h_{{\rm QNM},lm} \equiv \sum_{lm} h_{{\rm QNM}+,lm} + i h_{{\rm QNM}\times,lm} \equiv \sum_{lmn} C_{lmn} e^{-i \omega_{lmn} (t-t^{\ast})} \theta(t-t^{\ast}),
\label{model_ringdown_time}
\end{equation}
where $C_{lmn}$ is the excitation coefficient and $t^{\ast}$ is the start time of ringdown. The step function in the model, $\theta (t-t^{\ast})$, is introduced by assuming that ringdown starts instantaneously at $t=t^{\ast}$ \cite{Finch:2021qph}. In this manuscript, we consider the fitting of GW spectrum with a fitting function \cite{Finch:2021qph}
\begin{equation}
\tilde{h}_{{\rm QNM},lm} (\omega) = \frac{1}{2 \pi} \int dt h_{{\rm QNM},lm} (t) e^{i\omega t} = \frac{i}{2\pi} \sum_{n=0}^{n_{\rm max}}  \frac{C_{lmn}}{\omega -\omega_{lmn}} e^{i\omega t^{\ast}},
\label{fourier_model}
\end{equation}
which is the Fourier transformed function of (\ref{model_ringdown_time}). The phase factor of $e^{i \omega t^{\ast}}$ in (\ref{fourier_model}) originates from the step function in (\ref{model_ringdown_time}) and leads to the oscillation pattern in the spectral amplitude.
\begin{figure}[t]
  \begin{center}
    \includegraphics[keepaspectratio=true,height=90mm]{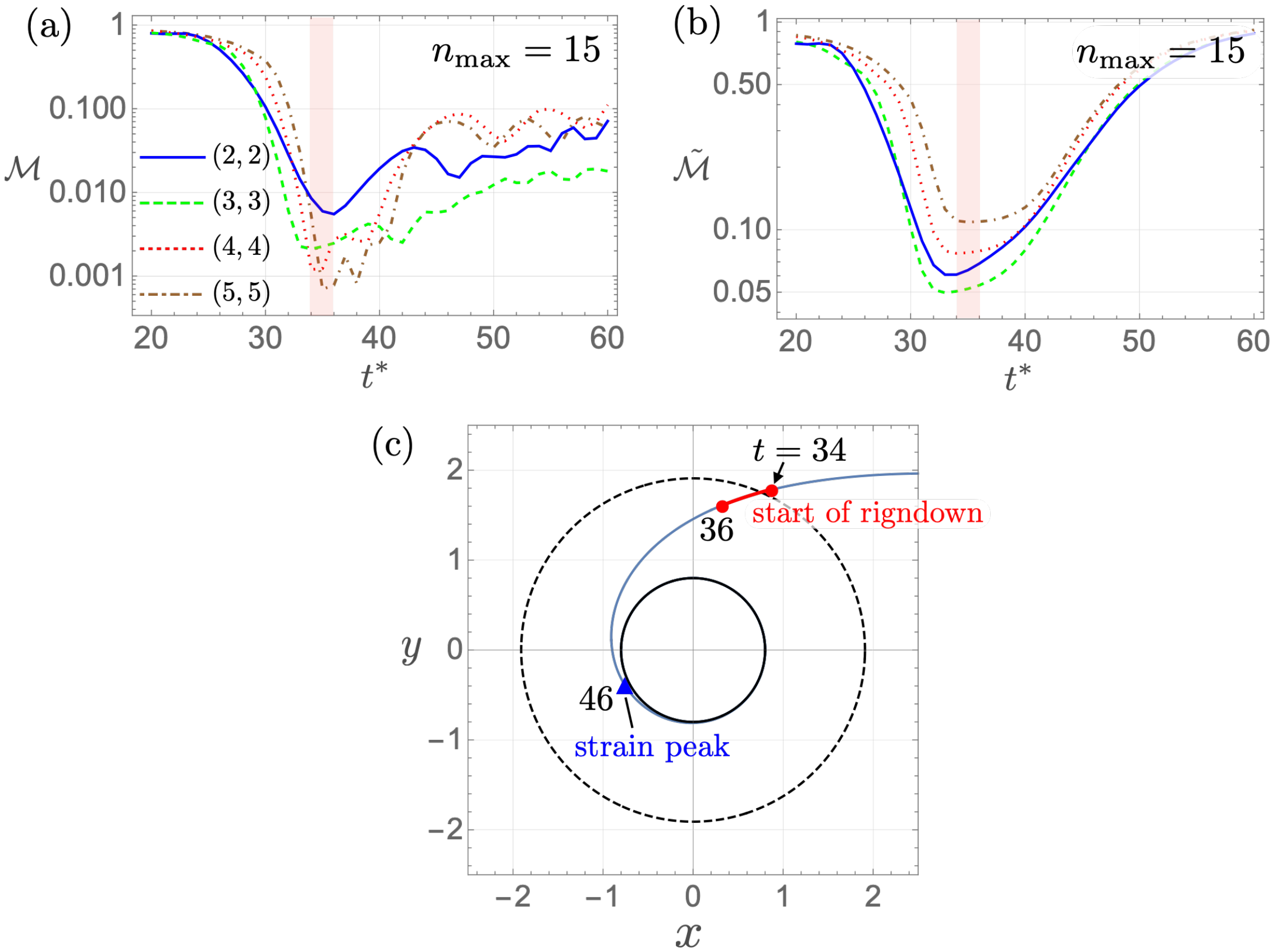}
  \end{center}
\caption{Mismatches, (a) ${\cal M}$ and (b) $\tilde{\cal M}$, for $j=0.8$ and $(l,m) = (2,2)$. We take into account the overtones up to $n = 15$. The bottom panel (c) shows the trajectory of the plungin particle along with the positions at which the GW signals, measured at $t= 34$ and $36$, are emitted (red circles). The uncertainty of the start time of ringdown, $34 \leq t \leq 36$ is shown with the red line. The blue triangle indicates the position at which the system emits the strain peak signal. The black solid and dashed lines indicate the outer horizon and photon sphere, respectively.
}
\label{mismatch}
\end{figure}

We then perform the fitting analysis in the frequency domain by using the unweighted least squares to determine $C_{lmn}$. The mismatch in the frequency domain is then obtained by
\begin{align}
\tilde{\cal M}_{lm} \equiv \left| 1- \frac{\braket{\tilde{h}_{lm}|\tilde{h}_{{\rm QNM},lm}}}{\sqrt{\braket{\tilde{h}_{lm}|\tilde{h}_{lm}} \braket{\tilde{h}_{{\rm QNM},lm} | \tilde{h}_{{\rm QNM},lm}}}} \right|,\\
\text{with} \ \ \braket{\tilde{f}|\tilde{g}} \equiv \int_{- \infty}^{\infty} d\omega \tilde{f}(\omega) \tilde{g}^{\ast} (\omega).
\end{align}
To see how well the fit in the frequency domain works, another mismatch is also computed by performing the inverse Fourier transform of $\tilde{h}$ and $\tilde{h}_{\rm QNM}$:
\begin{align}
&{\cal M}_{lm} \equiv \left| 1- \frac{\braket{h_{lm}|h_{{\rm QNM},lm}}}{\sqrt{\braket{h_{lm}|h_{lm}} \braket{h_{{\rm QNM},lm} | h_{{\rm QNM},lm}}}} \right|,\\
&\text{with} \ \braket{f|g} \equiv \int^{\infty}_{t^{\ast}} dt f(t) g^{\ast} (t).
\end{align}
In the following, we often omit the subscript of $l$ and $m$. The Figure \ref{mismatch} shows the mismatches, (a) ${\cal M}$ and (b) $\tilde{\cal M}$, with respect to $t^{\ast}$ and for several harmonics. The best-fit values of $t^{\ast}$ obtained for each angular mode\footnote{The definition of the best-fit $t^{\ast}$ in the manuscript is the value at which the mismatch, ${\cal M}$, takes the least value.} should be consistent with each other as $t^{\ast}$ is independent of angular modes. One can indeed read that ${\cal M}$ gives the best-fit value of $t^{\ast}$ in the range of $34 \lesssim t^{\ast} \lesssim 36$ for all angular modes we compute (Figure \ref{mismatch}-(a)). Also, the range of the best-fit values is well consistent with the result of $\tilde{\cal M}$ (Figure \ref{mismatch}-(b)). Figure \ref{mismatch}-(c) implies that the ringdown emission starts when the particle passes the photon sphere whose radius is $r= 2M \left\{1+\cos{[(2/3)\arccos{j}}]\right\}$. It is consistent with that the GW ringdown emission is governed by the photon sphere. On the other hand, the strain peak of the GW signal is emitted when the particle reaches the vicinity of the outer horizon.

\begin{figure}[t]
  \begin{center}
    \includegraphics[keepaspectratio=true,height=45mm]{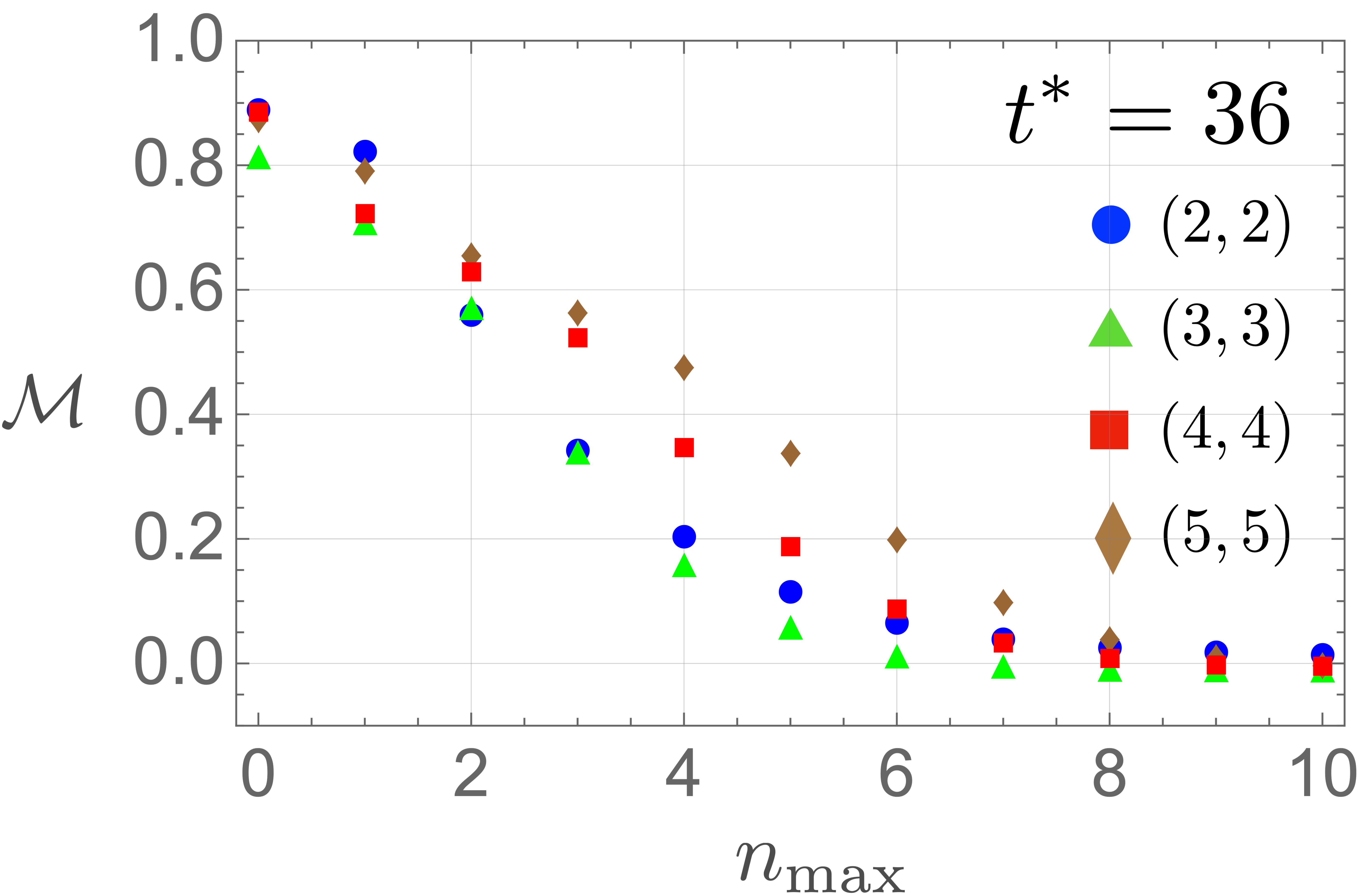}
  \end{center}
\caption{Mismatch, ${\cal M}$, with respect to $n_{\rm max}$. The spin parameter is $j=0.8$ and the start time is fixed to $t^{\ast} = 36$.
}
\label{convergence_nmax}
\end{figure}

To see the convergence of the QN-mode fitting with respect to the number of overtones included in the modeling, we compute the mismatch ${\cal M}$ with respect to $n_{\rm max}$ and the result is shown in Figure \ref{convergence_nmax}.
One can read that for higher harmonics, ${\cal M}$ converges at higher values of $n_{\rm max}$, but at less than $n_{\rm max} = 10$. It implies that the higher an angular mode is, the more overtones are excited\footnote{As the GW signal of $(l,m) = (2,2)$ includes a louder orbital signal, it makes the convergence rate of ${\cal M}$ with respect to $n_{\rm max}$ slow and the converged value of ${\cal M}$ for $l=m=2$ is the largest among computed angular modes.}. This can be analytically verified by computing the excitation factors \cite{Berti:2006wq,Zhang:2013ksa,Oshita:2021iyn} for higher harmonics and up to higher overtones as the excitation factor quantifies the {\it ease of excitation} of each QN mode\footnote{The most important overtone that has the highest value of the excitation factor is shown in Ref. \cite{Oshita:2021iyn} by computing the excitation factor up to the 20th overtone for various spin parameters.} independent of the source of a GW ringdown, and we will come back to this in our future work.
\begin{figure}[b]
  \begin{center}
    \includegraphics[keepaspectratio=true,height=130mm]{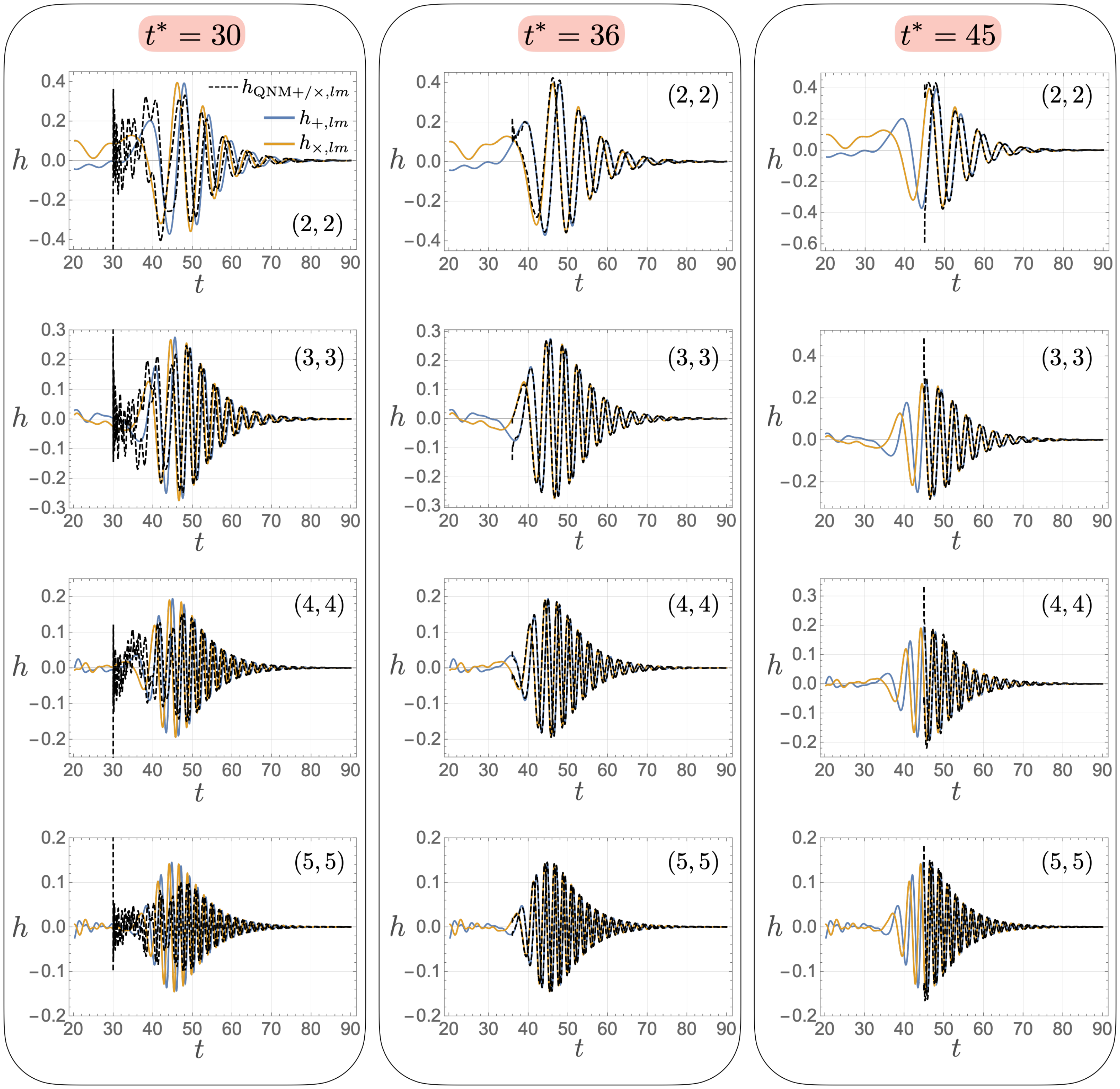}
  \end{center}
\caption{Result of the frequency-domain fitting analysis for $j=0.8$. Model functions, $h_{{\rm QNM}+/\times,lm}$, fitted to the GW signals with $n_{\rm max} = 15$ are shown for $l=m=2$, $3$, $4$, and $5$ with different three values of the start time of ringdown: $t^{\ast} = 30$, $36$, and $45$. The best-fit value of $t^{\ast}$ is $t^{\ast} \simeq 36$ and its results are shown in the center column.
}
\label{fit_time_domain_for_three}
\end{figure}

\begin{figure}[t]
  \begin{center}
    \includegraphics[keepaspectratio=true,height=70mm]{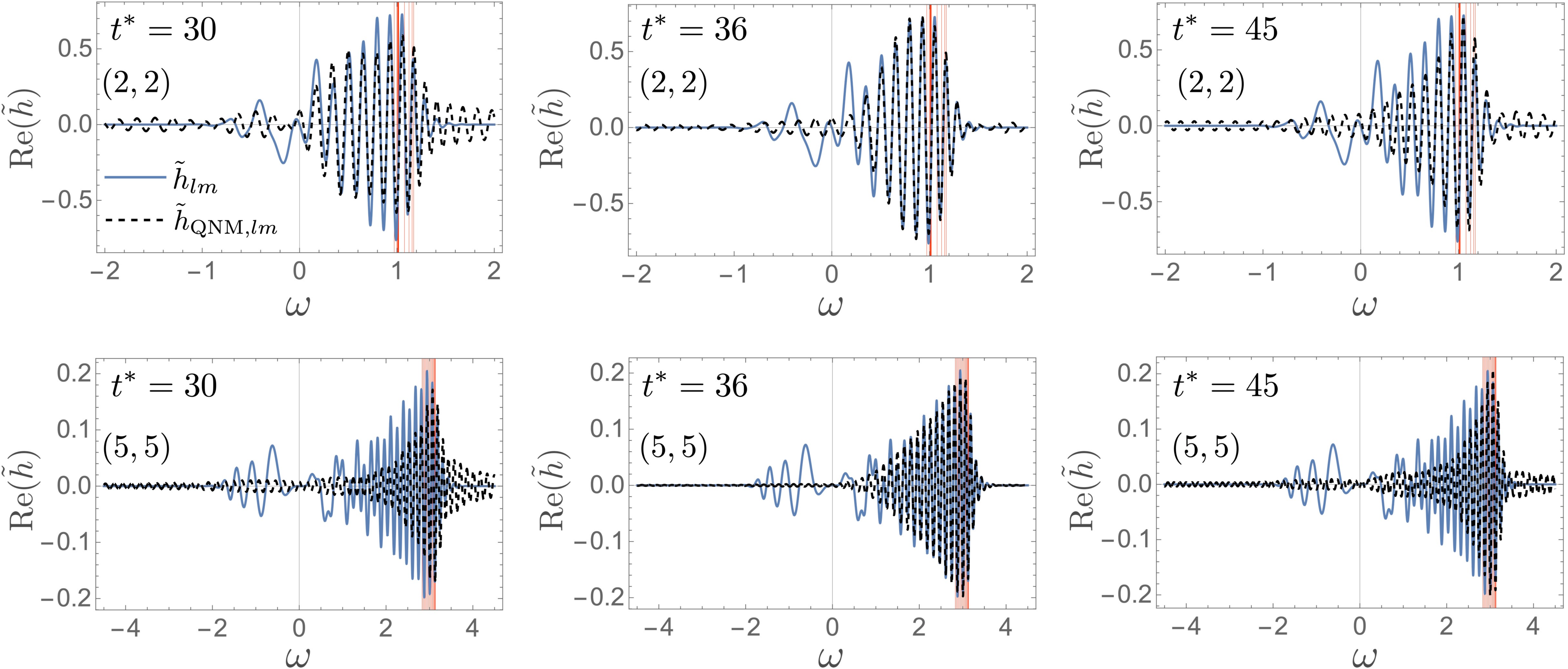}
  \end{center}
\caption{Result of the fitting in the frequency domain for the GW data of $j=0.8$. The angular modes are set to $(l,m)=(2,2)$ (upper) and $(5,5)$ (bottom). We set $t^{\ast} = 30$, $36$, and $45$. Red lines indicate the values of $\text{Re} (\omega_{lmn})$ from $n=0$ to $n=15$.
}
\label{interference}
\end{figure}

Figure \ref{fit_time_domain_for_three} shows the time-domain waveforms, $h$ and $\tilde{h}_{\rm QNM}$, obtained in the frequency-domain fit with $t^{\ast} = 30$, $36$, and $45$. One can see that the model waveform for $t^{\ast} = 30$ has noise at earlier times and that for $t^{\ast} = 45$ has a spiky noise at $t=t^{\ast}$. Those noise are visible in the frequency domain as is shown in Figure \ref{interference}. For $t^{\ast} = 30$ or $45$, the model function $\tilde{h}_{{\rm QNM},lm}$ (red dashed) has noise spreading over the long-range frequency domain. On the other hand, our analysis indeed works for a preferred value of $t^{\ast} = 36$. Also, one can see that the frequency-domain analysis is very efficient for higher harmonics (see the bottom panels in Figure \ref{interference}). This is because the orbital signal of $l=m>2$ emitted by a particle falling into the black hole is less significant compared to the ringdown signal. Therefore, the contamination from the orbital signal in $\tilde{h} (\omega)$ is smaller for higher harmonics.

\begin{figure}[t]
  \begin{center}
    \includegraphics[keepaspectratio=true,height=70mm]{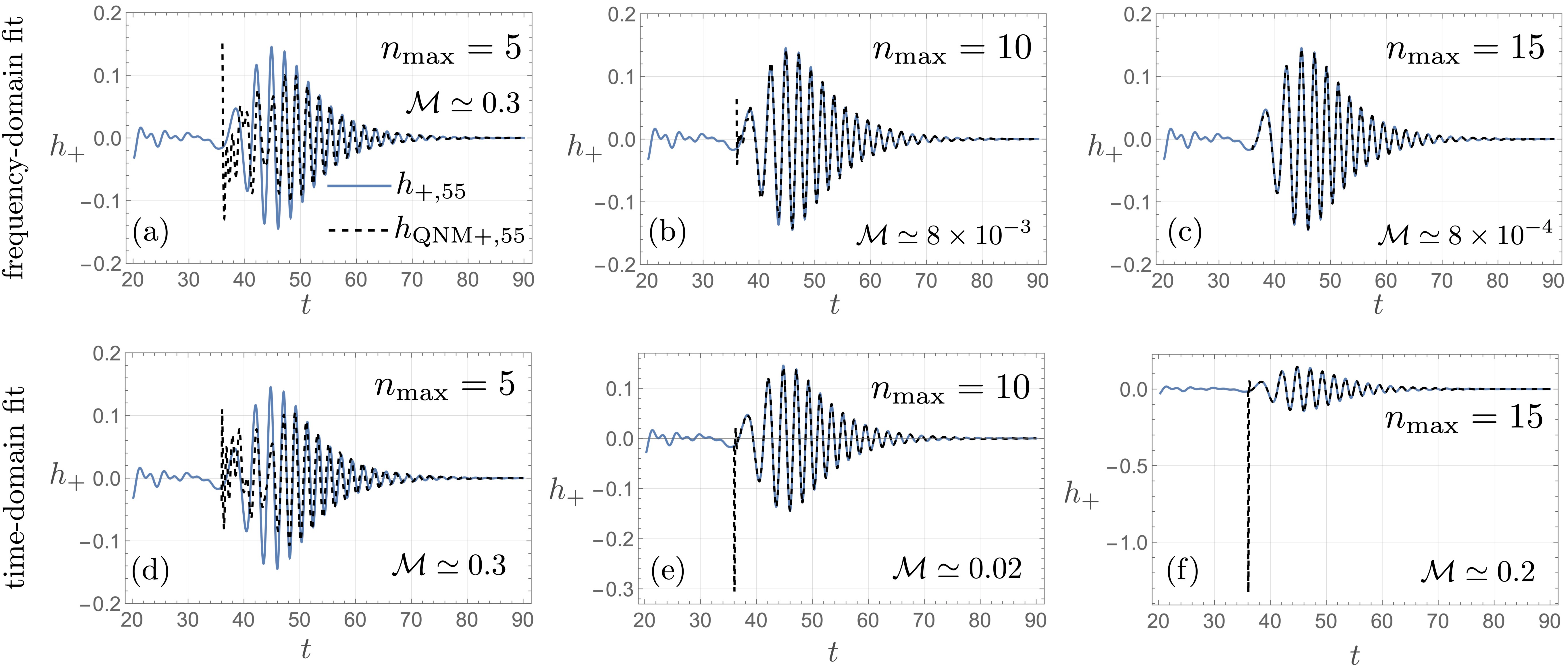}
  \end{center}
\caption{Results of the two types of fit are shown for the GW data of $j=0.8$ and $(l,m)=(5,5)$. These results show the convergence of the fit with respect to the inclusion of higher overtones in the frequency-domain (a, b, c) and time-domain fitting analysis (d, e, f). We perform the least-square fit in the frequency and time domains with the Mathematica's function \texttt{Fit}. We employ the Tikhonov regularization with the ridge parameter of $10^{-40}$. The result is insensitive to the exponent of the ridge parameter as long as it is much smaller than unity.
}
\label{FD_TD_comparison}
\end{figure}

The fit of QN modes in the frequency domain has the advantage of searching for the start time of ringdown, especially when it starts before the strain peak. The beginning of ringdown before the peak would involve a destructive interference among overtones, whose amplitudes at earlier times are exponentially amplified. As such, the fitting analysis in the time domain can be unstable against adding higher overtones since it requires controlling exponentially amplified overtones at earlier times (see Figure \ref{FD_TD_comparison}). On the other hand, in the frequency-domain analysis, we fit the Fourier transformed QN modes, $\sim e^{i \omega t^{\ast}}/(\omega - \omega_{lmn})$, which no longer have the exponential behaviour and does not lead to such an instability, as shown in Figure \ref{FD_TD_comparison}.

In summary, we confirm that the ringdown induced by a particle plunging into the rotating hole starts before the strain peak based on the frequency-domain analysis. This is caused by the destructive interference among the fundamental mode and overtones at earlier times. A significant destructive interference is possible when superposed modes have frequencies close to others, like the QN frequencies of a Kerr black hole. This would be supporting evidence that multiple overtones are simultaneously excited at earlier times. In the next section, we analyze the spectral data we numerically obtained and show that the excitation of multiple QN modes apparently follows the Fermi-Dirac distribution. We then propose a conjecture that the degrees of freedom of free oscillation of the Kerr black hole corresponds to that of a many-body thermal Fermi-Dirac system. It would be relevant to the Kerr/CFT correspondence, in which the agreement between the near-extremal QN modes and the Matsubara modes of the retarded Green's function in the corresponding CFT was demonstrated \cite{Guica:2008mu,Bredberg:2009pv,Chen:2010ni}.

\section{Thermal Excitation of Overtones and the Fermi-Dirac Statistics}
\label{sec_thermal_overtones}
\begin{figure}[t]
  \begin{center}
    \includegraphics[keepaspectratio=true,height=50mm]{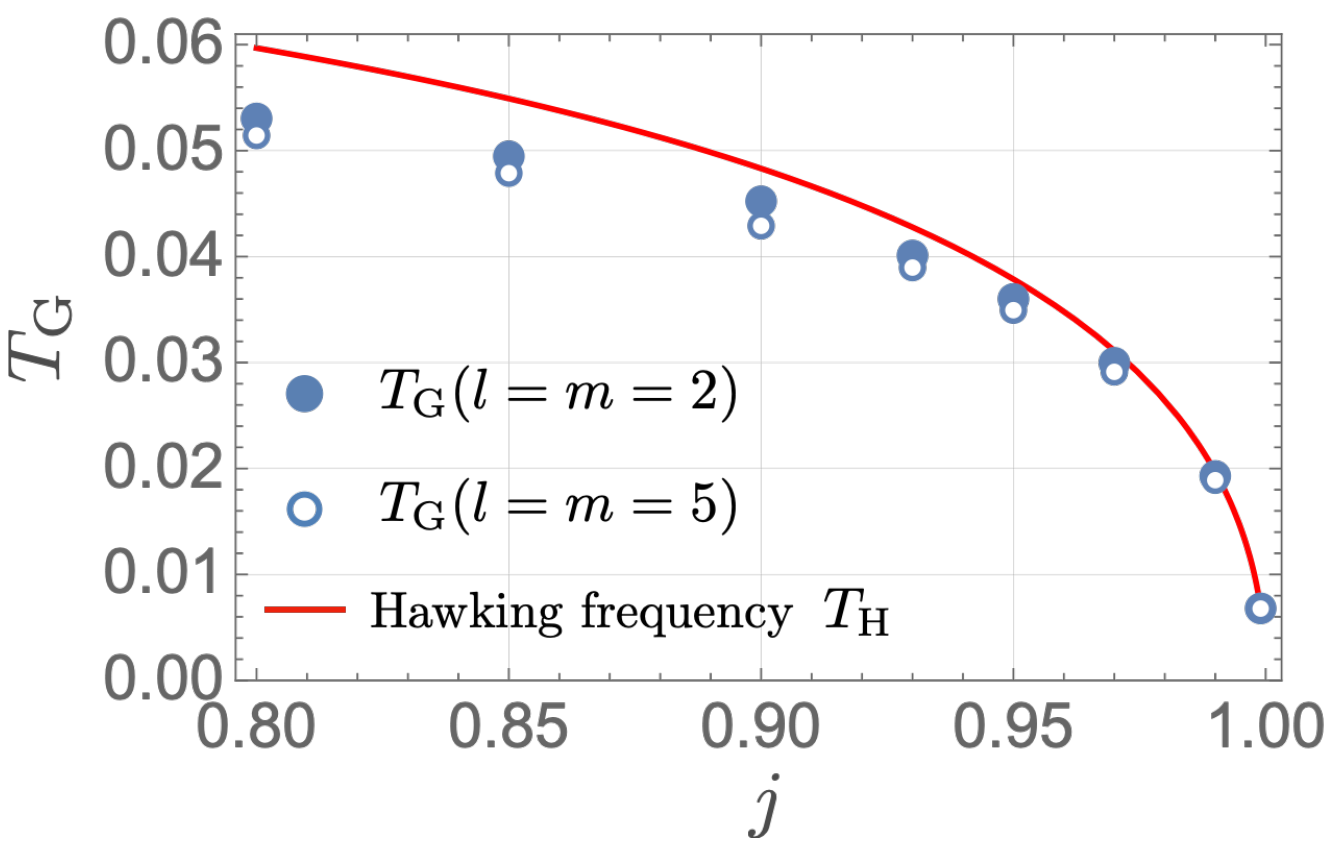}
  \end{center}
\caption{$T_{\rm G}$ for (filled circles) $l=m=2$ and (open circles) $l=m=5$. The red line shows the plot of the Hawking frequency $T_{\rm H}$. Best-fit values of $T_{\rm G}$ is obtained by fitting the Boltzmann factor, $e^{-(\omega - \mu_{rm G})/T_{\rm G}}$, to the spectral data of $|\tilde{h}_{lm}|^2$.
}
\label{greybody_F}
\end{figure}

In this section, we propose that the absolute square of the spectral amplitude, $|\tilde{h}_{+/\times,lm} (\omega)|^2$, can be modeled by the Fermi-Dirac distribution\footnote{As a qualitative statement, the square of the spectral amplitude can be regarded as {\it the number of particles} in the field theory.}
\begin{equation}
|\tilde{h}_{+/\times, lm}|^2 \sim \frac{1}{1 + e^{(\omega-\mu)/T}},
\label{FD_distribution}
\end{equation}
where $\mu$ and $T$ is constant with the dimension of frequency.
The plunging particle emits GW propagating to the angular momentum barrier and being scattered back to infinity, which leads to the emission of GW ringdown. Thus the reflectivity of the angular momentum barrier, $|{\cal R}_{lm}|^2$, may be imprinted on the signal, and one could read the greybody factor from the observed GW ringdown. The scattered part of $X_{lm}^{\rm (out)}$ in (\ref{X_out}) is
\begin{align}
X_{lm}^{\rm (scat)}(\omega) \simeq \frac{A_{lm} (\omega)}{2 i \omega B_{lm} (\omega)} \int_{r^{\rm (out)}}^{\infty} dr' \tilde{T}_{lm} (r', \omega) e^{i \omega r^{\ast}(r')}
+ \frac{1}{2 i \omega} \int_{r^{\rm (out)}}^{\infty} dr' \tilde{T}_{lm} (r', \omega) e^{-i \omega r^{\ast}(r')},
\label{scattered}
\end{align}
where $r^{\rm (out)} \gtrsim r^{\rm (light)}$ and $r^{\rm (light)}$ is the typical radius of the light ring. The first term in (\ref{scattered}) contributes to the ringdown signal as the factor of $1/B_{lm}(\omega)$ has the QN poles in the complex $\omega$ plane. On the other hand, the second term is irrelevant to the ringdown as it does not have the QN poles of $1/B_{lm}$.
The first term includes the factor of $A_{lm}/B_{lm}$ which is the reflective coefficient in the SN variable, and the reflectivity in energy flux is
\begin{equation}
|{\cal R}_{lm}|^2 = \frac{|C|^2}{|c_0|^2} \left| \frac{A_{lm}}{B_{lm}} \right|^2,
\end{equation}
where $|C|^2$ and $|c_0|^2$ are
\begin{align}
|C|^2 &\equiv \lambda^4 + 4 \lambda^3 + \lambda^2 (-40 a^2 \omega^2 + 40 am\omega + 4) + 48 a \lambda \omega (a \omega + m) +144 \omega^2 a^2 ( a^2 \omega^2 -2am \omega +m^2) + 144 \omega^2 M^2,\\
c_0 &\equiv \lambda (\lambda + 2) -12 a \omega (a \omega -m) -i12 \omega M.
\end{align}
As such, the absolute square of the GW spectrum of ringdown $\propto |X_{lm}^{\rm (scat)}|^2$ may include the greybody factor involving the Boltzmann suppression at higher frequencies\footnote{The factors of $|C|^2/|c_0|^2$ does not include the Boltzmann suppression, and as such, the exponential cut-off of the greybody factor originates from $|A_{lm}/B_{lm}|^2$.}. It can be modeled by the Fermi-Dirac distribution\footnote{This is the case only when the angular momentum potential is real \cite{Iyer:1986np,Konoplya:2019hlu}. Although the Teukolsky and Sasaki-Nakamura equation have the complex potentials, the Chandrasekhar-Detweiler equation has the purely real potential barrier \cite{Chandrasekhar:1976zz}.} \cite{Iyer:1986np,Konoplya:2019hlu}
\begin{equation}
|{\cal R}_{lm}|^2 \simeq \frac{1}{1+ e^{(\omega - \mu_{\rm G})/T_{\rm G}}},
\end{equation}
where $\mu_{\rm G}$ and $T_{\rm G}$ are constants\footnote{Note that the amplification factor is absent in this expression as the WKB approximation does not work when the two asymptotic frequencies, $\omega$ and $k$, are of the opposite sign \cite{Konoplya:2019hlu}.
For our purpose, i.e., modeling GW ringdown with the Fermi-Dirac-type function, this does not cause any problems as the Boltzmann suppression at higher frequencies is important to extract $T$ and $\mu$.}. The constant $\mu_{\rm G}$ takes a value close to the superradiant frequency $\mu_{\rm H} \equiv m \Omega_{\rm H}$ and $T_{\rm G}$ approaches to the value of the Hawking frequency $T_{\rm H} \equiv \sqrt{1-j^2}/(4 \pi r_+)$ in the limit of $j \to 1$ (Figure \ref{greybody_F}).

We indeed find that for near-extremal rotations, the spectrum of GW signal we numerically computed exhibits the Fermi-Dirac distribution in (\ref{FD_distribution}) with around the Hawking frequency $T_{\rm H}$ and the chemical potential of superradiant frequency $\mu_{\rm H}$:
\begin{equation}
T \simeq T_{\rm H}, \ \mu \simeq \mu_{\rm H}.
\end{equation}
We also find that for medium spins ($j \lesssim 0.9$) and for lower angular harmonics, the temperature and chemical potential take a value close to
\begin{equation}
T \simeq T_0 \equiv |\text{Im} (\omega_{lm0})|/ \pi, \ \mu \simeq \mu_0 \equiv \text{Re} (\omega_{lm0}).
\end{equation}
Based on these supporting evidence shown below in more detail, we conjecture that the degrees of freedom of a black hole ringing might follow the Fermi-Dirac statistics. Indeed, the QN frequencies of a Kerr black hole is very similar to the Matsubara frequencies of the Fermi-Dirac distribution (\ref{FD_distribution}).

\begin{figure}[t]
  \begin{center}
    \includegraphics[keepaspectratio=true,height=40mm]{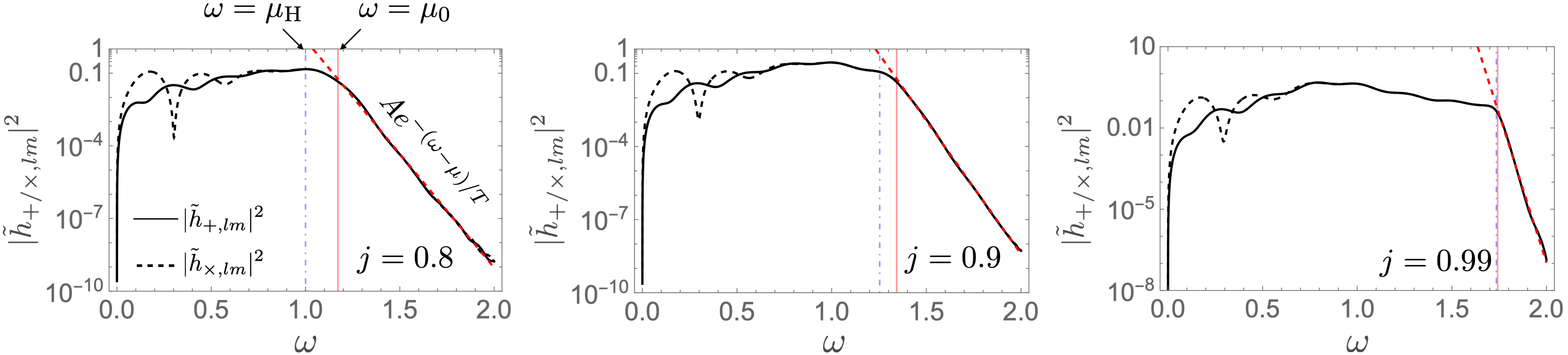}
  \end{center}
\caption{The absolute square of the spectral amplitude of the GW signals for $j=0.8$, $j=0.9$, and $j=0.99$ with $l=m=2$. The Boltzmann distribution (fitted with the red dashed line) appears at higher frequencies than $\omega = \mu_0$ (red solid line). The blue dot-dashed line indicates $\omega = \mu_{\rm H}$.}
\label{short_range}
\end{figure}

From the absolute square of spectral amplitude shown in Figure \ref{short_range}, the Boltzmann distribution is observed in the range of $\omega \geq \mu_0$ rather than in $\omega \geq \mu_{\rm H}$ although $\mu_0 \to \mu_{\rm H}$ in the extremal limit \cite{Hod:2008zz}. In the limit, the thermal excitation is suppressed, and a would-be Fermi degeneracy can be seen in frequencies lower than the superradiant frequency $\omega \leq \mu_{\rm H}$ (see the right panel in Figure \ref{short_range}). We fit the Boltzmann distribution, $A e^{-(\omega- \mu)/T}$, to the GW spectral data, where $\mu$ is fixed to $\mu = \mu_0$, and $A$ and $T$ are fitting parameters. Also, we use the GW data of $\omega \geq 1.1 \times \text{Re}(\omega_{lm0})$ in the fitting analysis for intermediate rotations ($j=0.8$ and $0.9$). For a near-extremal situation ($j=0.99$), we use the data of $\omega \geq 1.0 \times \text{Re}(\omega_{lm0})$. After the fitting analysis, we evaluate the following two relative errors of the best-fit $T$:
\begin{equation}
\Delta_0 \equiv |T-T_0|/T_0, \ \Delta_{\rm H} \equiv |T-T_{\rm H}|/T_{\rm H}.
\end{equation}
As a result (see Table \ref{table_temp}), we find that the best-fit value of $T$ is very close to $T_{\rm H}$ for the near-extremal case ($j=0.99$). This is consistent with the fact that QN modes for a near-extremal case have the separation of $2 \pi T_{ \rm H}$. On the other hand, the best-fit value of $T$ for medium spins and lower harmonics is closer to $T_0$ than $T_{\rm H}$. For higher harmonics, it is interesting to see the consistency with the Lyapunov exponent associated with an instability of null geodesics at the photon sphere, which is not our focus and left for future work.
\begin{table}[H]
\begin{center}
\begin{tabular}{c|ccc|ccc|ccc}
\firsthline
&\multicolumn{3}{c|}{$j=0.8 \ (T_{\rm H} \simeq 0.0597)$}&\multicolumn{3}{c|}{$j=0.9 \ (T_{\rm H} \simeq 0.0483$)}&\multicolumn{3}{c}{$j=0.99 \ (T_{\rm H} \simeq 0.0197)$} \\
\cline{2-10}
~$(l,m)$~    & $T$ & $\Delta_0$& $\Delta_{\rm H}$ &  $T$ & $\Delta_0$& $\Delta_{\rm H}$&   $T$ & $\Delta_0$& $\Delta_{\rm H}$  \\
\hline
$(2,2)$       & 0.0462(2) & {\bf4}\% & 23\% & 0.0397(2) & {\bf3}\%& 18\%&  0.0198(2) & {\bf6}\% & {\bf0.6}\% \\
$(3,3)$       & 0.0493(2)& {\bf0.6}\% & 17\% & 0.0375(1) & 10\% & 22\% & 0.0196(3) & {\bf5}\%& {\bf0.4}\%\\
$(4,4)$      & 0.0565(1) & 14\% & {\bf5}\% & 0.0454(2) & {\bf8}\%& {\bf6}\% & 0.0200(3) & {\bf6}\%& {\bf 2}\% \\
$(5,5)$      & 0.0552(2) & 10\% & {\bf8}\% &  0.0483(1) & 14\% & {\bf 0.03}\% & 0.0196(4) & {\bf 4}\% & {\bf 0.4}\%\\
\lasthline
\end{tabular}
\end{center}
\caption{The best-fit temperature, $T$, obtained by fitting the Boltzmann distribution, $A e^{-(\omega - \mu_0)/T}$, to the absolute square of the GW spectra with $j=0.8$, $0.9$, and $0.99$. The standard error of $T$ obtained with \texttt{NonlinearModelFit} in Mathematica is shown. The standard error less than 10\% are shown in bold.}
\label{table_temp}
\end{table}

\begin{figure}[t]
  \begin{center}
    \includegraphics[keepaspectratio=true,height=37mm]{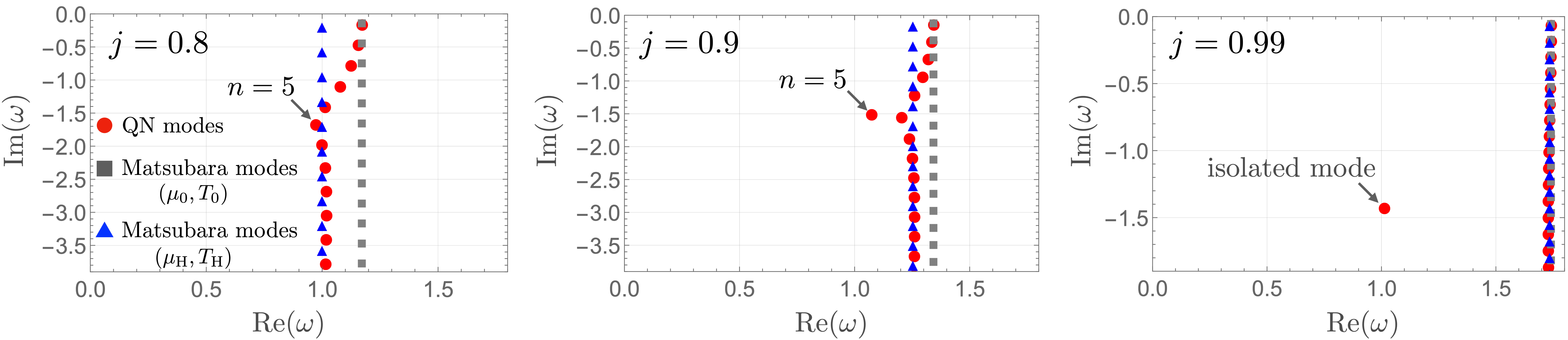}
  \end{center}
\caption{QN frequencies $\omega = \omega_{22n}$ (red circles) and the Matsubara frequencies $\omega = \omega_{{\rm M},n}$ for $(\mu_0, T_0)$ (gray squares) and $(\mu_{\rm H}, T_{\rm H})$ (blue triangles) with $l=m=2$ are shown. The spin parameters are set to $j=0.8$, $0.9$, and $0.99$.}
\label{qnm_Matsubara}
\end{figure}

Our conjecture is supported by the analogy between the QN modes of a near-extremal black hole and the Matsubara modes of the Fermi-Dirac distribution. Remember that the Fermi-Dirac distribution with a chemical potential, $\mu$, and temperature, $T$, has the Matsubara frequencies
\begin{equation}
\omega_{{\rm M}, n} (\mu ,T) = \mu -i \left( \frac{1}{2} + n \right) 2 \pi T,
\label{matsubara_poles}
\end{equation}
and indeed, the QN frequencies of a Kerr black hole can be approximated by the Matsubara frequencies (\ref{matsubara_poles}) especially for a near-extremal case as is shown in Figure \ref{qnm_Matsubara}. In the near-extremal limit, the imaginary part of QN modes indeed approach a half-integer times $2 \pi T_{\rm H}$ \cite{Hod:2008zz,Yang:2012pj}. Therefore, the apparently thermal excitation observed in the GW spectra may be understood as the excitation of multiple overtones.

In the context of the Kerr/CFT correspondence, the QN modes of a near-extremal black hole can be derived from the dual CFT \cite{Guica:2008mu,Bredberg:2009pv,Chen:2010ni}. The CFT absorption cross section is
\begin{equation}
\sigma \sim T_L^{2h_L-1} T_{R}^{2h_R -1} \left|\Gamma\left( h_L + \frac{i \bar{\omega}_L}{2 \pi T_L} \right) \right|^2 \left|\Gamma\left( h_R + \frac{i \bar{\omega}_R}{2 \pi T_R} \right) \right|^2,
\end{equation}
with
\begin{equation}
\bar{\omega}_L \equiv \omega_L -q_L \Omega_L, \bar{\omega}_R \equiv \omega_R -q_R \Omega_R,
\end{equation}
where $(h_L, h_R)$ are operator dimensions, $(q_L,q_R)$ are charges, $(T_L, T_R)$ are temperature, and $(\Omega_L, \Omega_R)$ are chemical potentials of the two-dimensional CFT.
The identification \cite{Guica:2008mu,Chen:2010ni}
\begin{equation}
T_R \equiv T_{\rm H}, \ \Omega_{R} \equiv \Omega_{\rm H}, \ h_R \equiv \frac{1}{2} -i \delta, \ \delta^2 \equiv 2m^2 -\frac{1}{4} - \lambda_{lm}, \ h_L = h_R - s, \ \omega_R \equiv \omega, \ q_R \equiv m, \bar{\omega}_L \equiv m,
\end{equation}
leads to the poles of $\sigma$ at
\begin{equation}
\omega \simeq m \Omega_{\rm H} - i \left( \frac{1}{2} + n \right) 2 \pi T_{\rm H},
\label{Kerr_qnm_extremal_CFT}
\end{equation}
where the spin $s = h_R-h_L$ takes $|s|=1$ or $2$ for the gauge field or the graviton, respectively \cite{Chen:2010ni}. Note that the left sector is associated with the azimuthal rotation and is not relevant to the QN modes. The poles (\ref{Kerr_qnm_extremal_CFT}) are consistent with the Kerr QN frequencies in the extremal limit \cite{Hod:2008zz}.
In this section, we do not extract the overtones from the GW signal, but nevertheless one can see the thermal excitation of overtones in the spectral amplitude in this way, i.e., the fit of the Boltzmann distribution to a GW spectrum in higher frequencies. It is a novel observable footprint of the thermal or holographic nature of a rotating black hole.

\begin{figure}[t]
  \begin{center}
    \includegraphics[keepaspectratio=true,height=35mm]{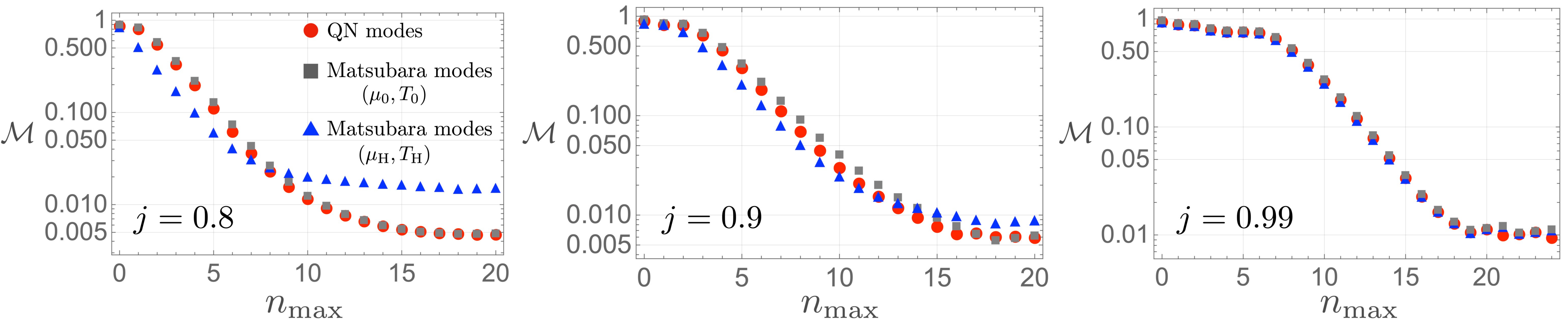}
  \end{center}
\caption{Mismatches, ${\cal M}$, computed with the model functions $\tilde{h}_{{\rm QNM},lm} (\omega)$ (red circles), $\tilde{h}_{{\rm M},lm} (\omega, \mu_0, T_0)$ (gray squares), and $\tilde{h}_{{\rm M},lm} (\omega, \mu_{\rm H}, T_{\rm H})$ (blue triangles). The angular mode is $l=m=2$.}
\label{convergence}
\end{figure}

To see how the ringdown modeling with the Matsubara modes works, let us fit the Matsubara modes to the GW data in the frequency domain as was done in the previous section. It is totally non-trivial whether the replacement of the QN-mode basis by the Matsubara-mode one works as a set of damped sinusoids cannot construct a complete basis in general. Nevertheless, we find that the mismatch, ${\cal M}$, evaluated with the Matsubara-mode modeling
\begin{equation}
\tilde{h}_{{\rm M},22} (\omega, \mu , T) = \frac{i}{2 \pi} \sum_{n=0}^{n_{\rm max}} \frac{C_{22n}}{\omega - \omega_{{\rm M},n} (\mu, T)}e^{i \omega t^{\ast}},
\label{fourier_model_matsubara}
\end{equation}
with $(\mu,T) = (\mu_0, T_0)$ agrees with the mismatch obtained by the QN-mode modeling for each value of $n_{\rm max}$ (Figure \ref{convergence}). It implies that the Matsubara modes in (\ref{matsubara_poles}) can be a proper basis to model the GW ringdown including the excitation of overtones although the Matsubara frequencies agree with the QN frequencies at lower tones only ($n \lesssim 2$).  On the other hand, the Matsubara-mode modeling with $(\mu,T) = (\mu_{\rm H}, T_{\rm H})$ works in the near-extremal limit only.

One can see that the mismatch reaches $0.01$ at around $n_{\rm max} \sim 11$, $14$, and $19$ for $j=0.8$, $0.9$, and $0.99$, respectively. It implies that the more rapid rotation the ringing black hole has, the more overtones are excited.
Also, we confirm that from our fitting analysis, the isolated mode of $j=0.99$ (see the right panel in Figure \ref{qnm_Matsubara}) does not contribute to the fit, which is consistent with the author's previous work on the excitation factors \cite{Oshita:2021iyn}, where the author confirmed that the absolute value of the excitation factor of the isolated mode is strongly suppressed. As such, we exclude the isolated mode from the fitting analysis shown in Figure \ref{convergence}.

\section{Discussions and Conclusions}
\label{sec_conclusions}
In this paper, we have studied the ringdown GW emission out of a Kerr black hole induced by a particle plunging into the hole. Then we have confirmed (i) the destructive interference of overtones before the strain peak and (ii) the apparent thermal excitation of multiple QN modes that can be modeled by the Fermi-Dirac distribution. This may be caused by the scattering of GW at the angular momentum barrier by which the greybody factor is imprinted on the ringdown spectrum. The apparent thermal nature of a black hole ringing can be tested by searching for the Boltzmann distribution in a GW spectrum at higher frequencies $\omega \gtrsim \text{Re} (\omega_{lm0})$.

Fitting multiple QN modes to GW waveforms in the frequency domain, we found that the ringdown starts before the strain peak and the significant destructive interference occurs as the real frequencies of overtones are close to that of the fundamental mode. Remember that the beating phenomenon, involving a destructive interference, is caused by the superposition of multiple sinusoids whose frequencies are close to each other. The fitting in the frequency domain works in a stable manner as the exponential amplification of the fitting mode functions is absent in the frequency domain. Also, one can perform the fitting analysis without truncating GW data beforehand and the start time of ringdown, $t^{\ast}$, can be handled as one of the fitting parameters. 

We have also found the Boltzmann distribution in the GW spectrum at higher frequencies sourced by an extreme mass ratio merger. The ringdown emission is induced by the light ring, and therefore, GW scattered at the light ring is observed as GW ringdown for a distant observer. The greybody factor that quantifies the scattering nature of the black hole may be imprinted on the ringdown. We have numerically confirmed that the ringdown can be modeled by the Fermi-Dirac distribution with the chemical potential of around the superradiant frequency and with the temperature of around the Hawking frequency for the Kerr black hole. This is reasonable as the greybody factor can be modeled by the Fermi-Dirac-type function based on the WKB approximation \cite{Iyer:1986np,Konoplya:2019hlu}. As the greybody factor is a no-hair quantity, i.e., it depends only on the mass and spin of a Kerr black hole, it may be possible to test general relativity by extracting the greybody factor, instead of extracting the QN frequencies, from the ringdown induced by a small mass ratio merger involving a large black hole.
A QN mode has the peak at $\omega = \text{Re} (\omega_{lmn})$ in the frequency domain since the spectral amplitude is $\sim 1/|\omega - \omega_{lmn}|$. As $\text{Re} (\omega_{lmn}) \sim \mu_{\rm H}$, we would argue that the exponential cut-off in the GW spectrum of $\sim e^{- (\omega - \mu_{\rm H})/T}$ can be explained by the interference effect among the fundamental mode and overtones. This implies that the overtones significantly contribute to GW ringdown on which the greybody factor is imprinted.

The would-be Fermi degeneracy was observed in the near-extremal case.
The suggestive relation between the black hole ringdown and Fermi-Dirac distribution is consistent with the fact that the QN frequencies of a Kerr black hole is very similar to the fermionic Matsubara frequencies, especially in the near-extremal situation. We have fit the Matsubara modes to the GW spectrum and have shown that it works as well as the fit with the Kerr QN modes. We conclude that those are supporting evidence of the thermal excitation of overtones.

It also implies that ringdown GW signals would include a footprint of the holographic nature of Kerr black holes, i.e., the correspondence between a many-body Fermi system and the free oscillation of a Kerr black hole. A well-known proposal relevant to this is the Kerr/CFT correspondence \cite{Guica:2008mu}. As a supporting evidence of the Kerr/CFT, the cross section of a near-extremal black hole agrees with the dual CFT prediction. The Matsubara frequencies of the retarded Green's function of the dual CFT also agrees with the Kerr QN modes \cite{Bredberg:2009pv,Chen:2010ni}
\begin{equation}
\omega_{lmn} \simeq m \Omega_{\rm H} - i 2 \pi T_{\rm H} \left(\frac{1}{2} + n\right),
\end{equation}
where $(l,m)$ is the label of an angular mode, $n$ is an overtone number, $\Omega_{\rm H}$ is the horizon frequency, and $T_{\rm H}$ is the Hawking frequency. We here consider the perturbation of the gravitational field that is a spin-2 field and follows the Bose-Einstein statistics, but we found that the statistics of a black hole ringing apparently follows the Fermi-Dirac distribution. It implies that there might exist a supersymmetric relation between the degrees of freedom of gravitational field and those of a black hole ringing. The thermal or holographic nature of rotating black holes can be testable without extracting overtones from GW data, and searching for the Boltzmann distribution in the spectrum of ringdown may put it to the test. The existence of near-extremal black holes has been reported in, e.g., Ref. \cite{Reynolds:2020jwt} (for a theoretical modeling, see \cite{Bustamante_2019}) and the ringdown signals emitted from those holes would be observable in the next-generation GW detectors, e.g., Laser Interferometer Space Antenna (LISA).

As possible future works, it will be important to include the self force of the plunging particle to beyond the extreme mass ratio. Also, it is important to study how sensitive the Fermi-Dirac statistics of GW ringdown is with respect to the orbital angular momentum or the initial kinetic energy of the plunging particle. It is also interesting to study ringdown emission with other spin-fields, e.g., scalar, vector, or Fermi fields, to see if they exhibit the would-be Fermi degeneracy in the spectra of ringdown. The detectability of the excitation of overtones or Boltzmann distribution with LISA will be studied elsewhere. The measurement of the Boltzmann distribution at higher frequencies of ringdown GW spectra emitted by black holes would enable us to read their mass and angular momentum without extracting their QN frequencies out of GW data, which may be another interesting direction to test general relativity.

\begin{acknowledgements}
The author thanks Sinya Aoki, Shinji Mukohyama, and Takahiro Tanaka for their insightful remarks and comments, Daichi Tsuna for the discussion on supermassive black holes which may have a rapid rotation. The author is supported by the Special Postdoctoral Researcher (SPDR) Program at RIKEN, FY2021 Incentive Research Project at RIKEN, and Grant-in-Aid for Scientific Research (KAKENHI) project for FY 2021 (21K20371).
\end{acknowledgements}

\appendix
\section{Explicit forms of $F_{lm}$, $U_{lm}$, $\gamma_{lm}$ and $W_{lm}$}
\label{app_explicit_forms}
In this Appendix, we show the explicit form of the functions, $F_{lm}$, $U_{lm}$, $\gamma_{lm}$ and $W_{lm}$ in (\ref{SNeq}) and (\ref{source_factor_SN}) to make the paper self contained. For the brevity, we will omit the subscript of $l,m$ in the following. For the derivation of $(F,U,\gamma)$ and $W$, see the original works \cite{Sasaki:1981sx} and \cite{Kojima:1984cj}, respectively.
\begin{align}
F &\equiv \frac{\Delta}{r^2+a^2} \frac{\gamma'}{\gamma},\\
\gamma &\equiv \sum_{i=0}^{4} c_n r^{-n},
\end{align}
where the prime denotes a derivative with respect to $r$ and
\begin{align}
c_0 &\equiv-12 i M \omega + \lambda (\lambda +2) -12 a\omega (a \omega - m),\\
c_1 &\equiv8ia \left\{3 a \omega - \lambda (a \omega - m)\right\},\\
c_2 &\equiv-24 i Ma(a \omega -m)+12 a^2\left\{1 -2 (a \omega -m)^2 \right\},\\
c_3 &\equiv24ia^3 (a\omega -m)-24 a^2M,\\
c_4 &\equiv12 a^4.
\end{align}
$U$ has the form of
\begin{equation}
U\equiv \frac{\Delta}{(r^2+a^2)^2} (U_1 + U_2 + U_3),
\end{equation}
where
\begin{align}
U_1 &\equiv \frac{-K^2}{\Delta} + \lambda + 4 - \frac{4 \Delta'}{r} + \frac{8 \Delta}{r^2} \left\{ 1+ \frac{i a(a \omega -m)}{B} \right\},\\
U_2 &\equiv -2+ \frac{1}{r^2+ a^2} \left( r \Delta' -2 \Delta + \frac{3a^2 \Delta}{r^2+a^2} \right),\\
U_3 &\equiv - \frac{\gamma'}{\gamma} \left\{ -iK+ \Delta \left( \frac{r}{r^2+a^2} + \frac{A+B'}{B} \right) \right\},\\
A &\equiv 3iK' + \lambda + 6 \Delta r^{-2},\\
B & \equiv -2iK +\Delta' -4\Delta r^{-1},\\
K & \equiv (r^2+a^2)\omega -am.
\end{align}
The function $W$ is
\begin{equation}
W \equiv W_1 + W_2 + W_3,
\label{W_function}
\end{equation}
where
\begin{align}
\frac{1}{m_{\rm p}} W_1 &\equiv f_0 e^{i (\omega \tilde{V} -m \tilde{\phi})} + \int^{\infty}_r dr_1 f_1 e^{i (\omega \tilde{V} -m \tilde{\phi})} + i \omega \int^{\infty}_r dr_1 \int^{\infty}_{r_1} dr_2 f_2 e^{i (\omega \tilde{V} -m \tilde{\phi})},\\
\begin{split}
\frac{1}{m_{\rm p}} W_2 &\equiv - \frac{a-L_{\rm p}}{\omega} (S_1 + (a \omega -m) S_0) \left[ \frac{r^2}{r^2+a^2} e^{i (\omega \tilde{V} -m \tilde{\phi})} \right.\\
& \left.+ \int^{\infty}_r dr_1 \frac{r_1^2}{r_1^2 + a^2} e^{i (\omega \tilde{V} -m \tilde{\phi})} \left\{ \frac{2a^2}{r_1 (r_1^2 + a^2)} + i (a\omega -m) \frac{a-L_{\rm p}}{\sqrt{Q}} - \frac{iam}{\Delta} \left( 1-\frac{P}{\sqrt{Q}} \right) \right\} \right],
\end{split}
\\
\frac{1}{m_{\rm p}} W_3 &\equiv S_0 (a-L_{\rm p})^2 \left[ - \frac{1}{2} \frac{r^2}{\sqrt{Q}} e^{i (\omega \tilde{V} -m \tilde{\phi})}
- \int^{\infty}_r dr_1 \frac{r_1}{\sqrt{Q}} e^{i (\omega \tilde{V} -m \tilde{\phi})} + \int^{\infty}_r dr_1 \int^{\infty}_{r_1} dr_2 \frac{1}{\sqrt{Q}} e^{i (\omega \tilde{V} -m \tilde{\phi})} \right],
\end{align}
and
\begin{align}
f_0 &\equiv - \frac{1}{\omega^2} r^2 \hat{S} \frac{\sqrt{Q}}{(r^2+a^2)^2},\\
\begin{split}
f_1 &\equiv - \frac{1}{\omega^2} \frac{r^2 \sqrt{Q}}{(r^2+a^2)^2} \left[ (S_1 + (a \omega -m) S_0) \frac{ia}{r^2} \right.\\
&\left.+ \hat{S} \left\{ \frac{2 (a^2-r^2)}{r (r^2+a^2)} + \frac{1}{2} \frac{Q'}{Q} +i (a \omega -m) \frac{a -L_{\rm p}}{\sqrt{Q}} - \frac{iam}{\Delta} \left( 1- \frac{P}{\sqrt{Q}} \right) \right\} \right],
\end{split}
\\
\begin{split}
f_2 &\equiv \frac{1}{\omega^2} \frac{r^2 \sqrt{Q}}{(r^2+a^2) \Delta} \left( 1- \frac{P}{\sqrt{Q}} \right) \left[ (S_1 + (a \omega -m) S_0) \frac{ia}{r^2} \right.\\
&\left.+ \hat{S} \left\{ \frac{2a^2}{r (r^2+a^2)} + \frac{2r}{r^2+(L_{\rm p}-a)^2} - \frac{(P + \sqrt{Q})'}{P + \sqrt{Q}} + i (a\omega -m) \frac{a-L_{\rm p}}{\sqrt{Q}} - \frac{iam}{\Delta} \left( 1- \frac{P}{\sqrt{Q}} \right) \right\} \right],
\end{split}\\
\hat{S} &\equiv (a\omega -m-ia/r) (S_1+(a\omega -m) S_0) - \lambda S_0/2,\\
S_0 &\equiv {}_{-2} S_{lm} (\pi/2),\\
S_1 &\equiv \frac{d}{d \theta} {}_{-2} S_{lm} (\pi/2),\\
d \tilde{V} &\equiv dt + dr^{\ast},\\
d \tilde{\phi} &\equiv d\phi + \frac{a}{\Delta} dr.
\end{align}

\section{Resolution of the numerical computation}
\label{app_resolution}
In this Appendix, we describe the details of our numerical computation solving the Sasaki-Nakamura equation in the frequency domain. To reduce the computation time while keeping the accuracy high, we divide the frequency space into $N$ segments:
\begin{equation}
\omega_{n-1} \leq \omega \leq \omega_n \ \text{with} \ n=1,2,...,N,
\end{equation}
and we take a larger (smaller) spatial step size in a lower-frequency (higher-frequency) segments. Our computation is performed in each segment with the frequency step size of $\Delta \omega_n = (\omega_n - \omega_{n-1})/60$ and with the spatial step size of $\Delta r^{\ast}_n$. The computation of the source term $\tilde{T}_{lm}$ involves multiple integrals on the particle trajectory. We compute it by the quadrature method with the step size of $\Delta r^{\ast}_{W}$.

\begin{figure}[h]
  \begin{center}
    \includegraphics[keepaspectratio=true,height=50mm]{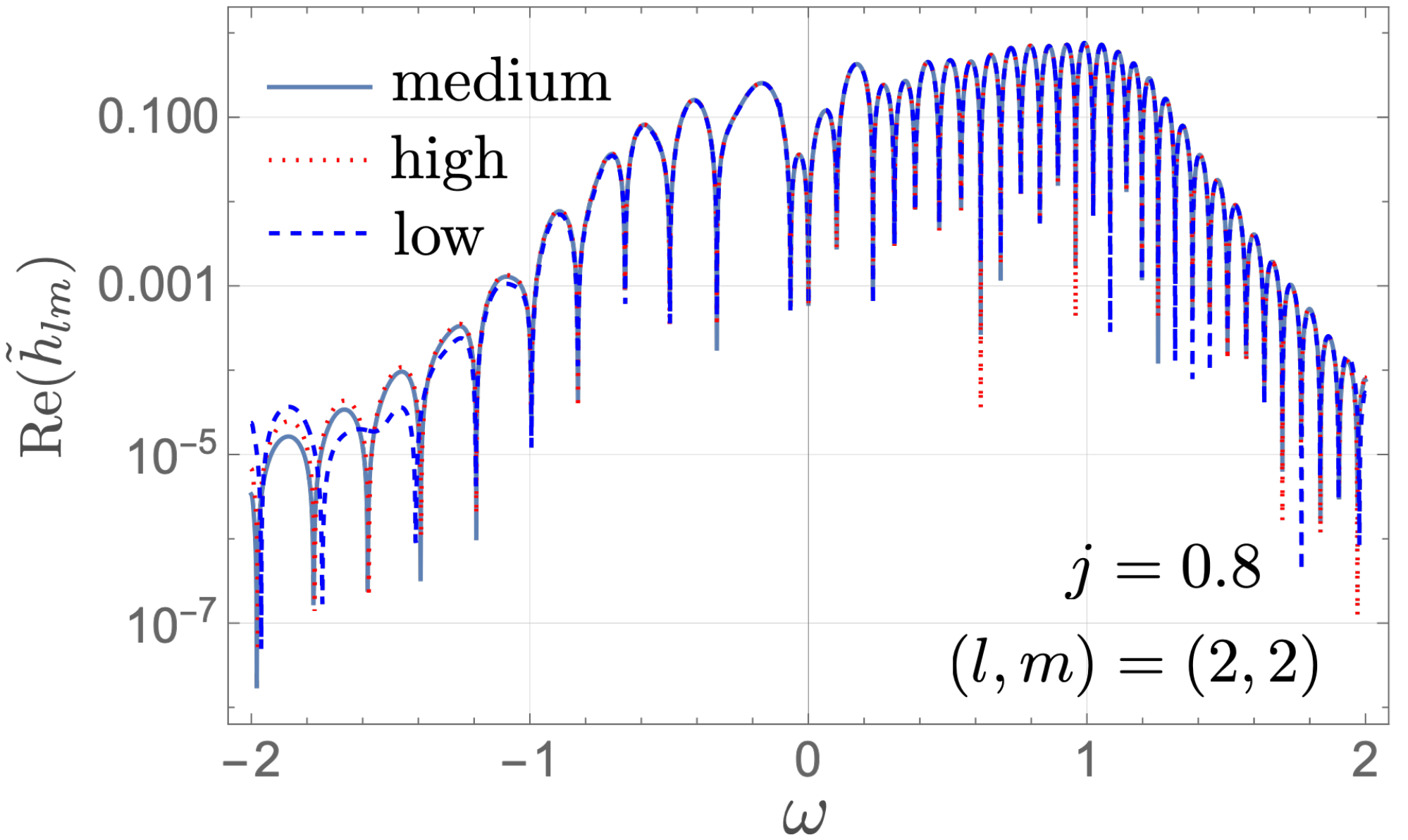}
  \end{center}
\caption{Resolution test for the numerical computation of GW spectrum with the medium resolution of $\Delta r^{\ast}_n$ (\ref{medium_reso}) and $\Delta r^{\ast}_W = 1/4$, low resolution (\ref{low_reso}), and high resolution (\ref{high_reso}).}
\label{resolution}
\end{figure}

Depending on a spin parameter and angular mode we consider, the resolutions and the frequency segments are optimized. As an example, let us consider the case of $j=0.8$ and $(l,m)=(2,2)$, where we take seven segments ($N=7$):
\begin{align}
\omega_0 = 1/200, \ \omega_1 = 1/10, \ \omega_2 = 1/2, \ \omega_3 = 3/4, \ \omega_4 = 1, \ \omega_5=5/4, \ \omega_6 = 3/2, \ \omega_7 = 2,
\end{align}
and we take the step size of
\begin{align}
\Delta r^{\ast}_n=
\begin{cases}
1/3 \ &n=1,\\
1/8 \ &n=2,\\
1/8 \ &n=3,\\
1/8 \ &n=4,\\
1/20 \ &n=5,\\
1/20 \ &n=6,\\
1/20 \ &n=7,
\label{medium_reso}
\end{cases}
\end{align}
and $\Delta r^{\ast}_W = 1/4$. To perform the resolution test, we perform the numerical computation with the above resolution and with another two resolutions:
\begin{align}
\Delta r^{\ast}_n \to 2 \times \Delta r^{\ast}_n, \ \Delta r^{\ast}_W = 1/2 \ &\text{(low resolution)},\label{low_reso}\\
\Delta r^{\ast}_n \to \frac{1}{2} \times \Delta r^{\ast}_n, \ \Delta r^{\ast}_W = 1/8 \ &\text{(high resolution)}.
\label{high_reso}
\end{align}
As is shown in Figure \ref{resolution}, the optimized (medium) resolution is high enough to compute the spectral peak at around $\omega \sim \text{Re} (\omega_{lm0})$ and to identify the Boltzmann distribution in the GW spectrum.

\section{Insensitivity of the thermal excitation of overtones to the orbit of the plunging particle}
\label{app_Lp}
\begin{figure}[H]
  \begin{center}
    \includegraphics[keepaspectratio=true,height=35mm]{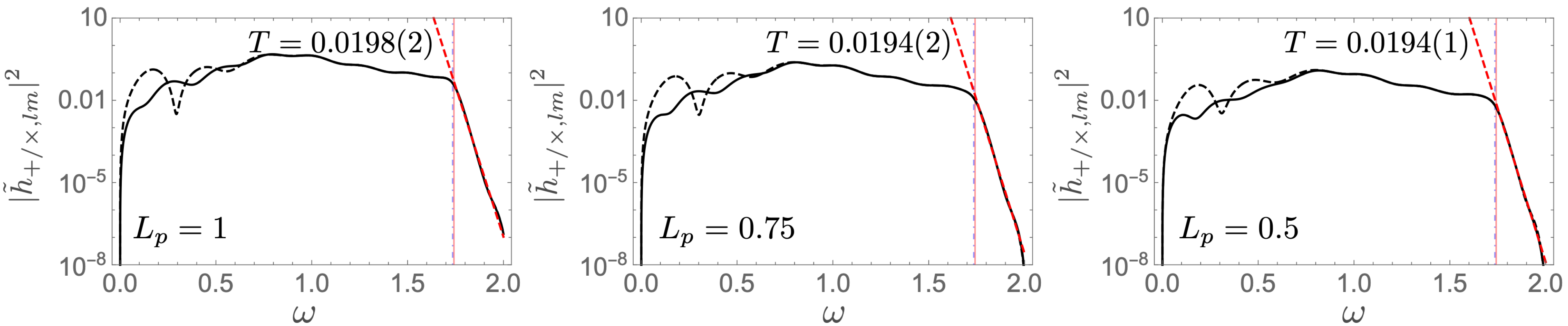}
  \end{center}
\caption{Absolute squares of the GW spectra for $j=0.99$, $l=m=2$. The orbital angular momentum is set to $L_{\rm p}=1$, $0.75$, and $0.5$. The Boltzmann factor is fitted to the data of $\omega \geq 1.0 \times \text{Re} (\omega_{220})$. The Hawking frequency for $j=0.99$ is $T_{\rm H} \simeq 0.197$. The relative error is $\Delta_{\rm H} \simeq 1 \%$ for $L_{\rm p}=0.75$ and $0.5$.}
\label{lp_check}
\end{figure}

In this Appendix, we show the insensitivity of the qualitative results presented in this paper to the orbital angular momentum of the plunging particle $L_{\rm p}$. We use $L_{\rm p}=1$ throughout the main text. Here we compute the GW spectra for $j=0.99$ and $(l,m)= (2,2)$ with $L_{\rm p}= 0.5$ and $0.75$. We then confirm that our results are insensitive to the value of $L_{\rm p}$. We check that the fit of the Boltzmann distribution leads to the best-fit temperature $T$ which is close to the Hawking frequency $T_{\rm H}$ even for $L_{\rm p}=0.5$ and $0.75$ (Figure \ref{lp_check}). The relative error to $T_{\rm H}$ is around $1\%$ for both cases. It implies that our main result of the thermal excitation of overtones is insensitive to the detail of the trajectory of the plunging particle. In this manuscript, we still fix the energy of the particle as $E= m_{\rm p}$, i.e., the particle is static at infinity, and restrict the trajectory on the equatorial plane ($\theta = \pi/2$). Relaxing these conditions makes the computation of the source term, $\tilde{T}_{lm}$, more complicated. A more detailed study on the thermality of ringdown for a more general plunging orbit is left for a future work.

%

\end{document}